# A GPU-Outperforming FPGA Accelerator Architecture for Binary Convolutional Neural Networks

YIXING LI, Arizona State University
ZICHUAN LIU, Nanyang Technological University
KAI XU, Arizona State University
HAO YU, Nanyang Technological University
FENGBO REN, Arizona State University

FPGA-based hardware accelerators for convolutional neural networks (CNNs) have obtained great attentions due to their higher energy efficiency than GPUs. However, it is challenging for FPGA-based solutions to achieve a higher throughput than GPU counterparts. In this paper, we demonstrate that FPGA acceleration can be a superior solution in terms of both throughput and energy efficiency when a CNN is trained with binary constraints on weights and activations. Specifically, we propose an optimized FPGA accelerator architecture tailored for bitwise convolution and normalization that features massive spatial parallelism with deep pipelines stages. A key advantage of the FPGA accelerator is that its performance is insensitive to data batch size, while the performance of GPU acceleration varies largely depending on the batch size of the data. Experiment results show that the proposed accelerator architecture for binary CNNs running on a Virtex-7 FPGA is 8.3x faster and 75x more energy-efficient than a Titan X GPU for processing online individual requests in small batch sizes. For processing static data in large batch sizes, the proposed solution is on a par with a Titan X GPU in terms of throughput while delivering 9.5x higher energy efficiency.

## 1 INTRODUCTION

Convolutional neural network (CNN) has become a popular machine learning engine for many image-related data analytics [15-16] [20] [27], such as image classification, face detection, object tracking, etc. CNNs outperform traditional feature selection based approaches especially for learning from big data. For a conventional CNN, high computation complexity and large memory footprint are the two main throughput bottlenecks for hardware acceleration. Therefore, the unmet throughput need of CNNs calls for the development of more efficient hardware acceleration solutions for driving real-time applications.

Several methods have been proposed to alleviate the computation complexity and memory footprint by reducing the redundancy of CNN models. These methods include pruning [18] [26], reduced-precision CNNs [4], and binary CNNs (BCNNs) [9]. The pruning technique [18] prunes the "useless" weights of a trained network based on sensitivity analysis, which can effectively reduce the CNN weight count (usually referred to as network size) for a ten-class classification problem by 75% [18]. Ref. 4 demonstrates that reducing the numerical precision of a CNN from 32 to 16 bits has very limited impact on classification accuracy. This can result in a network size reduction of 50%. However, a numerical precision below 8 bits resulted from quantization in the post-training stage often suffers from unacceptable accuracy drop [4]. Alternatively, recent advancement in binary-constrained deep learning has opened up new opportunities for efficient hardware acceleration. BinaryConnect [5] and the work in Ref. 6 demonstrate the successful use of binary and ternary (-1, 0, +1) weights in a CNN, respectively. But, they both have non-binary activations. As one step forward, EBP [7], Bitwise DNNs [8], and the BCNN in Ref. 9 successfully exploit both binary weights and activations. In particular, the BCNN in Ref. 9 shows a 0.96% classification error rate on the MNIST database [17], which is comparable to a full-precision state-of-the-art CNN. Overall, BCNNs have been shown with up to 96.8% reduced network sizes with minimum accuracy loss when comparing to their full-precision counterparts. Therefore, it is believed that BCNN is a more hardware-friendly model with superior accuracy-complexity trade-off.

Thus far, GPU-based CNN accelerator is still dominant due to its improved throughput over CPUs. However, the high power consumption of GPUs has brought up cooling concerns in data center computing. On the other hand, FPGA-based CNN accelerator has been widely investigated due to its energy efficiency benefits. As the system throughput is proportional to the computing parallelism and operating frequency, the theoretical throughput of GPU-based and FPGA-based CNN accelerators can be estimated on the 1st order based on device specifications. A Titan X GPU has 3,072 CUDA cores, while a Virtex-7 FPGA has 3,600 DSP48 slices. For implementing a full-precision CNN, the computing parallelism of GPUs and FPGAs can be approximately the same. But, GPUs offer 5-10x higher frequency. As a result, FPGAs can hardly match up the throughput of GPUs for accelerating full-precision CNNs. Differently, for a BCNN, the operations in the convolution layers become bitwise XNORs and bit-count logic. A direct impact is that one can use LUTs instead of DSP48 slices to implement the bitwise operations on an FPGA. Hundreds of thousands of LUTs make it possible for a high-end FPGA to match up or surpass the throughput of a GPU, even considering the bitwise operation capability of CUDA cores. Moreover, FPGAs benefit from much higher energy efficiency, which makes it a superior solution for accelerating BCNN in a data center setting. Early research effort [9] shows that GPU can get 7x speedup using a binary kernel for MNIST classification task on a binary multilayer perceptron (MLP). However, there have been very few studies on exploring FPGA-based accelerator architecture for binary neural networks.

In this paper, we propose an optimized FPGA accelerator architecture tailored for BCNN. The proposed architecture was adopted to implement a 9-layer BCNN on a Xilinx Virtex-7 XC7VX690 FPGA, which achieves nearly state-of-the-art classification accuracy on CIFAR-10. The experiment results show that the FPGA implementation outperforms its optimized GPU counterpart with 75x higher energy efficiency and 8.3x higher throughput for processing a small batch size of 16 images (e.g. from individual online request). For processing a large batch size of 512 images (e.g. from static data), the FPGA implementation achieves comparable throughput with 9.5x higher energy efficiency compared with the GPU counterpart.

The contributions of this paper are summarized as follows:

• We propose a throughput optimization model for the end-to-end mapping of general BCNNs.

• We demonstrate a 7.663-TOPS 8.2-W FPGA accelerator for a BCNN that highly outperforms the GPU counterpart especially for processing individual online requests in small batch size for the 1$^{st}$ time.

• We reveal the impact of applying binary constraints in CNN training on FPGA acceleration is the enablement of massive computing parallelism of bitwise operations based on abundant LUT resources.

• We optimize the accelerator architecture to fully exploit both spatial and temporal parallelism across all the layers using architectural unfolding, pipelining, and data-flow control with memory channels. Compared with GPU implementations that only have spatial parallelism, the proposed architecture offers superior throughput and energy efficiency performance regardless of the size of workload.

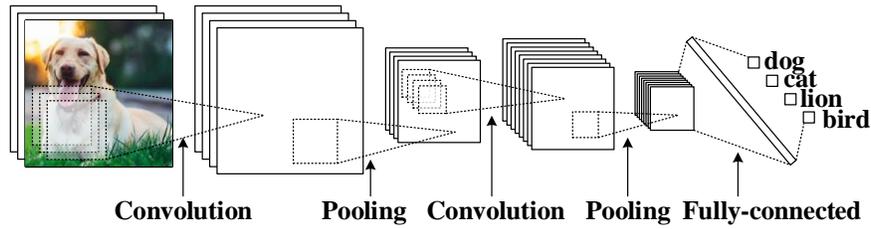

Fig. 1. Convolutional neural network.

## 2 BACKGROUND AND MOTIVATION

### 2.1 CNN

A CNN is a trained neural network model with high-level features extracted from input images [13]. A typical CNN model contains convolutional, pooling, and fully-connected layers as shown in Fig. 1. The first few layers usually capture regional information such as edges and curves, and the last few layers interpret these low-level features into high-level abstractions with the posterior probability assigned for classification.

*2.1.1 Convolution.* The convolution layer is the core layer of a CNN. Taking an RGB image as an example, the input of each convolutional layer is a 3D feature map with the size of $WID' \times HEI' \times DEP'$ as shown in Fig. 2. Each filter has a size of $FW \times FH \times FD$, where $FW$ and $FH$ is the width and height of the reception field, respectively, and $FD$ is equal to the depth $DEP'$ of the input feature maps. N filters are constructed as a 4D tensor. The output feature maps $Y$ in the size of $WID \times HEI \times DEP$ are obtained from the spatial convolution along the 1$^{st}$ and the 2$^{nd}$ dimensions of the input feature maps with the 3D-filter $W[n]$. The operation in

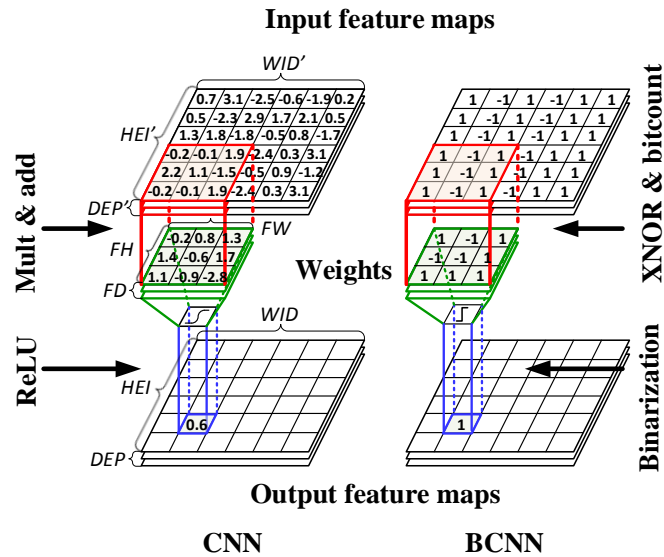

Fig. 2. A single layer in CNN and BCNN.

convolutional layers is defined as

$$Y[n][w'][h'] = \sum_{w=0}^{FW-1} \sum_{h=0}^{FH-1} \sum_{d=0}^{FD-1} W[n][w][h][d] \times fmap[w'+w][h'+h][d]. \quad (1)$$

One should note that there is no data dependency for the calculation of each pixel across the entire output feature maps. Therefore, spatial parallelism can be applied in the hardware architecture to improve throughput. Differently, within the convolution operation for calculating each pixel, data dependency exists among the nested loops of summation in (1). These data-dependent operations can be unfolded and pipelined in the hardware architecture to gain temporal parallelism and improve throughput.

*2.1.2 Pooling.* The pooling layer performs subsampling across a K×K contiguous region on the output feature map of convolutional layers. Pooling is used to pool out sensitive information critical to classification and eliminate insensitive information that is irrelevant. Also, pooling layers reduce an amount of trainable parameters in the network. There are two kinds of pooling methods which are commonly used in CNNs. One is max-pooling, which picks the maximum value of the pooling region. The other is average-pooling, which picks the mean value of the pooling region.

*2.1.3 Normalization.* Normalization is a powerful technique that stabilizes and accelerates the training process [11]. In the inference stage, normalization needs to be applied to match the training process. Statistical reference values are counted across the whole training set as

$$z = \frac{y - \mu}{\sqrt{\sigma^2 + \epsilon}} \gamma + \beta, \quad (2)$$

where $\mu$ is the mean value, and $\sigma^2$ is the variance with very a small constant $\epsilon$ to ensure a non-zero denominator. Note that $\gamma$ and $\beta$ scales and shifts the normalized values, respectively. Since $\mu, \sigma^2, \epsilon, \gamma$ and $\beta$ are all constants in the inference stage, they can be precomputed to reduce the computation complexity of normalization.

*2.1.4 Nonlinear function.* Nonlinear function is an element-wise operation that performs on each neuron after the normalization in the convolutional layers and the fully-connected layers. Two common nonlinear functions used in CNNs are Sigmoid and Rectified Linear Unit (ReLU) [13].

## 2.2 Binary CNN (BCNN)

A BCNN is a CNN trained with binary constraints that results in binary weights and activations, and a significant reduction in computation complexity. The convolution operation is the most time-consuming and computation-intensive part of a CNN. In a BCNN, as shown in Fig. 2, both the weights and activations are constrained to a binary set of values, e.g. [+1, -1]. As such, the multiplications in convolution is simplified to a bitwise exclusive NOR (XNOR). From a vector operation perspective, the convolution can be expressed as an XNOR dot-product operation as

$$Y[n][w'][h'] = \sum_{w=0}^{FW-1} \sum_{h=0}^{FH-1} \sum_{d=0}^{FD-1} \overline{W}[n][w][h][d] \oplus \overline{fmap}[w'+w][h'+h][d]. \quad (3)$$

Comparing to a real-valued CNN with a single–precision data format, the FPGA implementation of a BCNN requires much reduced logic and memory resources. Although, one should note that neither the inputs nor the outputs of the normalization and the pooling layers are binarized. The BCNN adopts a max-pooling scheme, which is thought to be more hardware-friendly than average-pooling [14]. Since the weights and activations are constrained to either +1 or -1, the nonlinear function becomes an adjusted sign function, a.k.a. a Binarize function defined as

$$Binarize(z) = \begin{cases} 1 & if\ z \geq 0, \\ 0 & otherwise. \end{cases} \quad (4)$$

**2.3 Compression ratio and accuracy of compact CNNs**

Table 1 shows some popular techniques for neural network compression. The baseline is a standard CNN trained by conventional techniques resulting in a full precision network for inference. Experiment results show that simply quantizing the network parameters below 10 bits in the post-training stage will cause significant accuracy drop on CIFAR-10 classification task using the CNN model in Ref. 9. Although, pruning the network has limited accuracy loss, the pruned network is still based on full-precision operations. The compression ratio achieved by pruning can be up to 5x [18], but the hardware resources needed for computing the remaining full-precision operations still have the same logic complexity.

Differently, the BCNN trained with binary constraints features the best compression ratio with superior accuracy performance. [9] shows that BCNN can achieve same accuracy as the full-precision CNN on a ten-class classification task on CIFAR-10 dataset. Ref. 19 demonstrates that with improved training technique, the BCNN only suffers from a 5% accuracy drop in terms of both top-1 and top-5 error for a 1000-class classification task based on ImageNet dataset. In addition, the hardware resources needed for realizing the bitwise convolutions in BCNNs are just simple logic gates rather than multipliers. All of these suggest that BCNNs offer much superior trade-off between complexity and accuracy and are ideal for efficient hardware implementation.

**2.4 Impact of binarization on hardware acceleration**

A Titan X GPU has 3,072 CUDA cores (one ALU per core) and can run at up to 1 GHz, while a midrange Virtex-7 FPGA has 3,600 DSP48 slices and 433,200 LUTs and typically runs at around 100-200 MHz. For mapping a full-precision or reduced-precision CNN, the two devices are on a par in terms of the level of computing parallelism considering that a CUDA core and a DSP48 slice can map a floating- and a fixed-point multiplication accumulator (MAC), respectively. But, FPGAs run at a 5-10x lower frequency in general. As a result, the existing FPGA implementations of reduced-precision CNNs can hardly achieve comparable throughput to their GPU counterparts.

A BCNN offers large room for throughput improvement for both GPU-based and FPGA-based implementations. When using a tailored binary kernel on a GPU, a fully-pipelined ALU in one CUDA core can process 32 bitwise operations per clock cycle. This increases the equivalent computing parallelism of a Titan X GPU to 3,072×32=98,304 for running a BCNN. On the other hand, for an FPGA-based BCNN, the bitwise operation can be efficiently mapped onto the abundant LUT resources. Since one 6-input LUT can map 2.5 XNORs on average, the computing parallelism of a Virtex-7 FPGA is on the order of 433,200×2.5≈1,000,000. Given the operation frequency difference, GPU- and FPGA-based BCNN implementations should have a similar level of throughput performance in a 1st order estimation. The FPGA-based solution features much higher energy-efficiency. It is also worth mentioning that GPUs can only achieve the theoretical peak throughput when the data batch size is large enough to hide the computation and memory access latency. Thus, in the application scenarios such as processing online classification requests from individual users where small batches of data must be processed on the fly, FPGA-based solution will keep the promise to outperform GPU counterparts in terms of both throughput and energy efficiency. In the following sections, we present an FPGA-based BCNN accelerator and a benchmarking study that validate our hypothesis.

Table 1. Methods for neural network compression

| Methods | Execution stage | Compression ratio | Inference | Accuracy |
|---|---|---|---|---|
| Standard | Training | 1x | full precision + full network | lossless |
| Quantizing | post-training | Up to 3x | reduced precision + full network | lossy |
| Pruning | training | Up to 5x | full precision + pruned network | lossless |
| BNN | training | Up to 32x | binary + full network | lossless |

### 2.5 A BCNN on CIFAR-10

In order to assess the practical performance of the proposed architecture, we use the BCNN on CIFAR-10 [9] as an example model for the FPGA implementation. The overall architecture of BCNN is shown in Table 2 [9]. It takes an RGB image with a size of 3 × 32 × 32 as the input of the first layer. For each convolutional layer, the filter size is fixed as 3 × 3 with a stride and zero padding of 1 pixel each. The filter specification of each convolutional layer in Table 2 is denoted as the *WID×HEI×DEP*. Max-pooling is performed over a 2 × 2 window with a stride size of 2 followed by the convolutional layers of 2, 4 and 6. The last three layers are fully connected layers. Normalization is applied to all the layers, which is followed by binarization except for the last layer.

## 3 ALGORITHM REFORMULATION FOR EFFICIENT FPGA MAPPING

### 3.1 Binary-encoded Convolution

When training the BCNN in [9], the weights and activations are constrained to either +1 or -1. For efficient FPGA mapping, we encode +1/-1 as 1/0 in our design. In this way, it only takes 1 bit to store a weight or an activation value. Moreover, the convolution operation in layer $l$ is simplified into an XNOR dot product of the input feature map $a_{l-1}^b$ and the weight $w_l^b$, given as

$$y_l = XnorDotProduct(a_{l-1}^b, w_l^b). \tag{5}$$

Equation (5) sums up 1s and 0s, which is different from the original BCNN that sums up -1s and +1s in (3). The relation between the original output feature map $y_{lo}$ and the revised $y_l$ in our

Table 2. BCNN configurations

| Name | CONV-1 | CONV-2 | CONV-3 | CONV-4 | CONV-5 |
|---|---|---|---|---|---|
| **Filter/weight** | 3×3×3 | 128×3×3 | 128×3×3 | 256×3×3 | 256×3×3 |
| **# of filters** | 128 | 128 | 256 | 256 | 512 |
| **Output size** | 128×32×32 | 128×16×16 | 256×16×16 | 256×8×8 | 512×8×8 |
| **Name** | CONV-6 | FC-1 | FC-2 | FC-3 | |
| **Filter/weight** | 512×3×3 | 8192×1024 | 1024×1024 | 1024×10 | |
| **# of filters** | 512 | - | - | - | |
| **Output size** | 512×4×4 | 1024 | 1024 | 10 | |

design can be expressed as

$$y_{lo} = 1 \times y_l + (-1) \times (cnum_l - y_l) = 2y_l - cnum_l, \quad (6)$$

where $cnum_l = FW \times FH \times DEP$ is the total number of bitwise XNOR operations needed for each $y_{lo}$. The difference between $y_{lo}$ and $y_l$ is compensated in the normalization module in our design.

Note that all the layers take the binary feature map of its previous layer as the input except for the first layer. In our design, we rescale the input data within the range of [-31,31] and use a 6-bit fixed-point data format, which helps to reduce the resource utilization of non-binary operations at the cost of a limited classification accuracy loss of <0.5%. Since the input image size is 3 × 32 × 32, the computational complexity of the first layer is not a dominating factor. The fixed-point dot product of a 6-bit signed input $a_0$ and a 2-bit signed weight $w_1$ is denoted as

$$y_1 = FpDotProduct(a_0, w_1). \quad (7)$$

### 3.2 Comparator-based Normalization

The parameters subject to training can be considered as constant values in the inference stage. Therefore, we can combine the binarization in (4), the normalization function in (2) and the value compensation in (6) into a modified sign function defined as

$$NormBinarize(y_l, c_l) = \begin{cases} 1 & if\ y_l \geq c_l, \\ 0 & otherwise, \end{cases} \quad (8)$$

where $c_l$ is a constant threshold derived by $c_l = (cnum_l + \mu - \beta\sqrt{\sigma^2 + \epsilon}/\gamma) \times 0.5$, and it is rounded to the nearest integer for hardware implementation.

The impact of the proposed reformulation on hardware implementation is that both the reformulated normalization and binarization functions can be efficiently implemented as a single LUT-based comparator. In addition, one only needs to store one threshold value $c_l$ for each output value rather than a set of training parameters $\mu, \sigma^2, \beta$ and $\gamma$.

```
{1. The first layer}
y₁ ← FpDotProduct(a₀, w₁)
a₁ ← NormBinarize(y₁, c₁)
{2. Remaining hidden layers}
for l = 2 to 8 do
        yₗ ← XnorDotProduct(a^b_{l-1}, w^b_l)
        If (l = 2,4,6) then
                yₗ ← MP(yₗ)
        end if
        aₗ ← NormBinarize(yₗ, cₗ)
end for
{3. Output layers}
yₗ ← XnorDotProduct(a^b_{l-1}, w^b_l)
aₗ ← Norm(yₗ, cₗ)
```

Fig. 3. Pseudo code of the BCNN algorithm.

### 3.3 BCNN Model Overview

We summarize the inference flow for the reformulated BCNN algorithm in Fig. 3. The convolution in the $1^{st}$ layer involves fixed-point dot product operations ($FpDotProduct$). Differently, bitwise XNOR dot product operations ($XnorDotProduct$) are used in all the other layers. Max-pooling ($MP$) is applied in layers 2, 4 and 6. Normalization and binarization are combined as a single function ($NormBinarize$), which is applied in all layers except for the output layer. The output layer ends with the normalization function $Norm$ for classification.

## 4 ARCHITECTURE DESIGN AND OPTIMIZATION

### 4.1 Architecture Overview

The binary nature of the BCNN enables us to map all the weights, feature maps, and reference values (for normalization) onto the on-chip block RAMs (BRAMs) in a single FPGA. This eliminates any DRAM access latency and dramatically reduces the energy consumption of the system comparing to the existing work relying on off-chip storage [1] [3] [12] [21].

Fig. 4 shows the overall architecture of the proposed BCNN accelerator. The binary convolutional kernel in each layer is followed by a NormBinarize (NB) kernel with or without a Max-pooling (MP) kernel. All of the kernels are highly parallelized with an optimized number of processing elements (PEs) and operate in a single instruction multiple data (SIMD) fashion. A streaming architecture is enabled by using double-buffering-based memory channels to handle the data flow between adjacent layers. Each PE in the binary convolutional kernel handles an XNOR dot product operation, which is the core operation in both convolutional and fully-connected layers. The PEs interface with the BRAMs in parallel to read the weights concurrently.

### 4.2 Architectural parameters

*4.2.1 Loop Unrolling.* Note that the three nested loops in (3) that accumulate the XNOR output values along the three dimensions of a convolutional filter has loop-carried data dependency. Unrolling data-dependent loops is the same as architectural unfolding, which will improve throughput by increasing the level of temporal parallelism. This trades off more hardware resource with improved computing parallelism. The unfolding factor is a critical architectural parameter in our design, denoted as *UF*. *UF* has a maximum value of $WID \times HEI \times DEP$ in each layer.

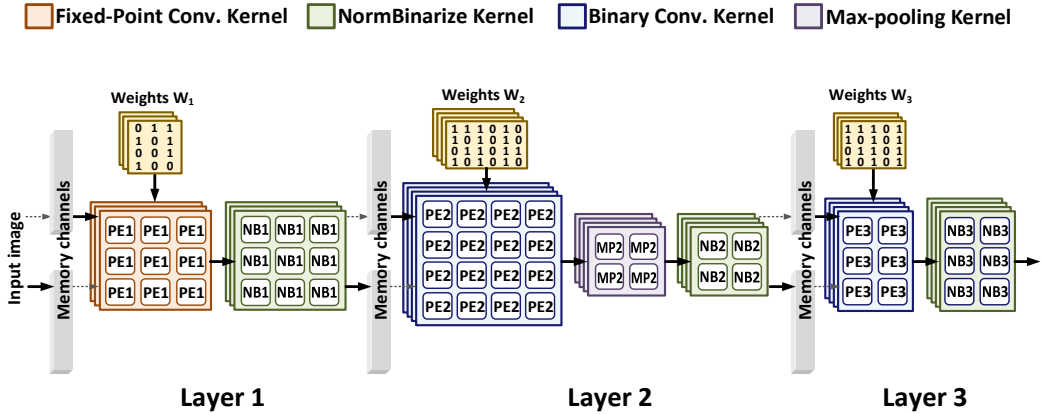

Fig. 4. Overview of the proposed accelerator architecture for BCNN.

Differently, the calculation of the pixel values along the three dimensions of an output feature map has no loop-carried data dependency. Unrolling independent loops is equivalent to creating spatial parallelism in the architecture to improve throughput. In our design, we fully unroll these independent loops to maximize the throughput. We denote the unrolling factor of independent loops as *P*. Maximizing *P* generates a massively parallelized PE array by utilizing the abundant LUT resources on the FPGA. Note that the PEs in the same layer are identical, but they could be different in size across different layers.

*4.2.2 Pipelining.* Loop pipelining is applied in the proposed architecture to further enhance the temporal parallelism and maximize the system throughput. Note that the queuing time to feed in the next data is the inversely proportional to throughput, which is referred to as initial interval *I* in this paper. If there is a loop existing in the data path, the minimum initial interval will be limited by the loop latency of the recursive architecture. With loop pipelining, we can feed in the next data whenever possible with the minimum initial interval. In the case of a fully pipelined implementation, we can feed in new data every clock cycle ($I = 1$).

## 4.3 Throughput Modeling and Optimization

If we only perform one XNOR operation and one accumulation in each clock cycle, the total execution time $Cycle_{conv}$ in terms of clock cycles of a convolutional layer can be model as

$$Cycle_{conv} = WID \times HEI \times DEP \times FW \times FH \times FD, \tag{9}$$

where *WID*, *HEI*, and *DEP* denotes the width, height, and depth of a convolutional filter, and *FW*, *FH*, and *FD* denotes the width, height and, depth of an output feature map, respectively.

When architectural unfolding is applied in performing the XNOR dot product operation in each PE, $Cycle_{conv}$ will be divided by *UF*. Similarly, when spatial parallelism is applied to create PE arrays for processing *P* output pixels in parallel, $Cycle_{conv}$ will be further reduced by *P* times. The same PE array is reused to calculate the output feature maps with pipelining applied, which contributes to an *I*-cycle initial interval for the most inner loop. Thus, the throughput of the convolutional kernel with architectural optimization can be formulated as

$$throughput_{CONV} = \frac{UF \times P}{Cycle_{conv}} \times \frac{1}{I} \times freq, \tag{10}$$

where $freq$ is the system frequency. Note that $throughput_{CONV}$ is inversely proportional to the estimated cycle count $Cycle_{est}$ in a convolutional layer, defined as

$$Cycle_{est} = \frac{Cycle_{conv}}{UF \times P} \times I. \tag{11}$$

In the proposed accelerator architecture, we use a double buffering scheme to further enhance the spatial parallelism of the system as shown in Fig. 4. The computation of each layer is triggered at the same time and alternates between two phases. Specifically, one channel of $fmap_{L-1}$ is used as the input of the $L^{th}$ layer while the $L$-$1^{th}$ layer is writing new outputs into the other $fmap_{L-1}$ channel. When both layers finish processing, the memory buffers swap, and the next processing phase is triggered. Therefore, the overall system-level throughput can be formulated as

$$throughput = \frac{max(C_1, C_2, C_3 \ldots, C_k)}{freq}, \quad (12)$$

where $C_L$ is the execution time of the $L^{th}$ layer in the proposed accelerator architecture. $C_L$ can be either $Cycle_{est}$ for throughput modeling or $Cycle_r$ for evaluating real execution throughput. One should note that the system throughput can be maximized with the optimal hardware utilization when all the layers have equal execution time ($C_1 = C_2 = C_3 = \cdots = C_k$). In the case that the $L^{th}$ layer has longer execution time than other layers, one can always increase the parallelism of the $L^{th}$ layer while decreasing that of other layers to gain throughput with minimum overhead in resource usage. Since the convolutional layers take up over 95% of the computation, we only emphasize the optimization of convolutional layers in this section. The fully-connected layer can be easily optimized to match up the system throughput using the same principle.

## 5 FPGA IMPLEMENTATION

In this section, we present the strategy of mapping different computing units to maximize the FPGA resource utilization.

### 5.1 PE Unit

The block diagram of a PE unit is shown in Fig. 5. A PE unit handles the XNOR dot product operation of a weight vector and a feature map vector from the previous layer. The vectors are fed into an array of 2-input XNOR gates followed by a parallelized bit-count logic for accumulation. Since both the XNOR gates and the bit-count logic take binary values as input, the PEs can be efficiently implemented using the abundant LUT resources. This is the key to enabling massive computing parallelism on an FPGA. Note that the number of XNOR gates in each PE is the same as the unfolding factor *UF* of the current layer. By accumulating the PE output, the pixel value of an output feature map can be computed by the bit-count logic.

### 5.2 Computing Kernels

Fig. 6 shows the architecture of the convolutional kernel followed by the Max-pooling and NormBinarize kernels. Each convolutional kernel has an array of PEs implemented using LUTs followed by an array of accumulators implemented using DSP48 slices. The number of PEs and DSP slices is equal to the spatial parallelism factor *P*. Each convolutional kernel thereby computes *P* pixel values of the output feature map in parallel. Besides the weight arrays, only intermediate results of the accumulator outputs (bit-count results) within a single feature map are stored in BRAMs. Feature maps are mapped onto distributed RAMs.

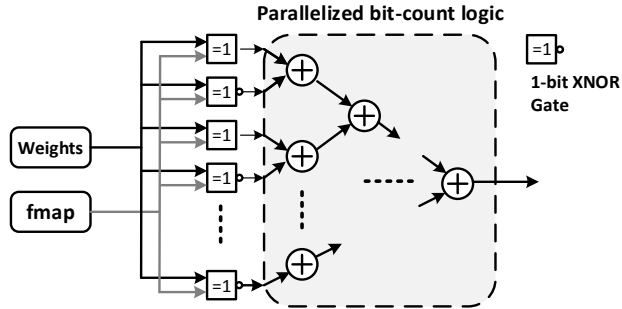

Fig. 5. Processing element (PE).

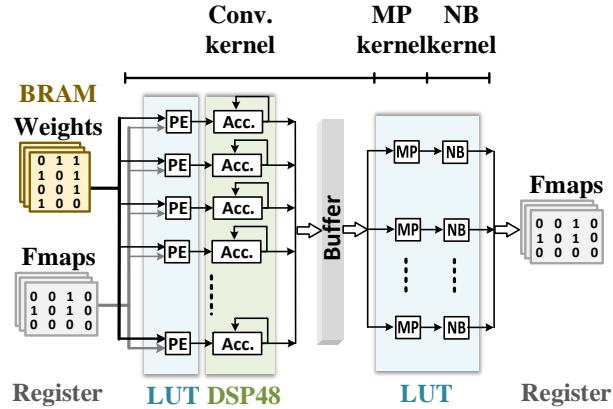

Fig. 6. The architecture of computing kernels and their FPGA mapping schemes.

For the convolutional layers 1, 3 and 5 without max-pooling, the outputs of accumulators are directly connected to the NB kernels. The hardware kernel of fully-connected layers is similar to Fig. 6. Note that the max-pooling is performed in pipeline with the computation of feature maps in our implementation.

### 5.3 Memory

To read and write a large number of bits in the same clock cycle, we have to partition and reshape the memory arrays in the BCNN model. Partition essentially breaks down a large data array into smaller ones to fit in multiple BRAMs for parallel access. Reshaping basically redefines the depth and width of a single BRAM by grouping multiple words into a wider one. In our design, the weight and $fmap$ arrays are mapped onto BRAMs and distributed RAMs (registers), respectively. Since the maximum word length of a BRAM in a Virtex-7 FPGA is limited to 32 bits, we first reshape the weight array by 32 and then partition the weight arrays into several BRAMs to guarantee enough memory bandwidth for the required system throughput.

## 6 EXPERIMENT RESULTS

Table 3. Optimized parameters for each layer

| Layer  | UF   | P  | $Cycle_{conv}$ | $Cycle_{est}$ | $Cycle_r$ |
|--------|------|----|----------------|---------------|-----------|
| Conv 1 | 27   | 32 | 3538944        | 4096          | 5233      |
| Conv 2 | 384  | 32 | 150994944      | 12288         | 12386     |
| Conv 3 | 384  | 16 | 75497472       | 12288         | 12296     |
| Conv 4 | 768  | 16 | 150994944      | 12288         | 13329     |
| Conv 5 | 768  | 8  | 75497472       | 12288         | 12386     |
| Conv 6 | 1536 | 8  | 150994944      | 12288         | 14473     |

Table 4. FPGA resource utilization summary

| Resource | LUTs | BRAMs | Registers | DSP |
|---|---|---|---|---|
| Used | 342126 | 1007 | 70769 | 1096 |
| Available | 433200 | 2060 | 607200 | 2800 |
| Utilization/% | 78.98 | 48.88 | 14.30 | 39.14 |

We implement the proposed accelerator architecture for the BCNN in Ref. 9 using the optimal architectural parameters shown in Table 3. We optimize the parameters of *UF* and *P* to make $Cycle_{est}$ of each layer approximately the same based on the throughput model in (12). Each layer is also fully pipelined with an initial interval of $I = 1$. Note that the operations along the *FW* and the *FD* dimensions are fully unfolded for maximizing the throughput.

### 6.1 Design Environment

We use C language to describe the accelerator architecture. Vivado HLS is used to produce the RTL codes. The Vivado Design Suite is used to map the design onto a Xilinx Virtex-7 XC7VX690 FPGA. The execution time in terms of clock cycles is reported by Vivado HLS and the system frequency is reported by Vivado Design Suite after the implementation stage. We notice a large discrepancy of LUTs usage between the synthesis reports in Vivado HLS and Vivado Design Suite. For accurate results, the resource utilization and power consumption are reported in Vivado Design Suite after the implementation stage.

### 6.2 FPGA Implementation results

As shown in Table 3, the real execution time $Cycle_r$ given by the synthesis report for each layer is well aligned with $Cycle_{est}$ estimated by our model in (11). The throughput bottleneck is layer 6 in this case. Running at a system frequency of 90 MHz, the FPGA-accelerated BCNN achieves an image processing throughput of 6,218 frames per second (FPS), which is the highest throughput for the same dataset reported by far. The top-1 accuracy rate is 87.8%, which is only 0.3% lower compared to the software model in Theano.

Table 5. Results in comparison with FPGA-based accelerators

| | Device | Clock (MHz) | Bit-width | GOPS | Power (W) | Energy Efficiency (GOPS/W) | Performance Density (GOPS/kLUT) |
|---|---|---|---|---|---|---|---|
| [3] | Virtex 6 | 200 | 16 | 147 | 10 | 14.7 | 0.98 |
| [1] | Virtex 7 | 100 | 32 float | 62 | 18.7 | 3.3 | 0.14 |
| [12] | Zynq-7000 | 150 | 16 | 137 | 9.6 | 14.3 | 0.75 |
| [4] | Stratix-V | 120 | 8 ~ 16 b | 117.8 | 25.8 | 4.56 | 0.45 |
| [22] | Arria-10 | 150 | 8 ~ 16 b | 645.25 | 21.2 | 30 | 4.01 |
| [23] | Intel QuickAssist QPI FPGA | 200 | 32 float | 123.48 | 13.18 | 9.37 | 0.62 |
| [24] | Arria-10 | 385 | fixed | 1790 | 37.46 | 47.78 | 4.19 |
| [21] | Zynq-7000 | 143 | 1 ~ 2 b | 207.8 | 4.7 | 44 | 4.43 |
| **Ours** | Virtex 7 | 90 | 1 | 7663 | 8.2 | 935 | 22.40 |

To reduce runtime, we adopt a bottom-up design strategy by synthesizing our design layer by layer in Vivado HLS and implementing the entire system in Vivado Design Suite. The overhead introduced by initialization is negligible. Table 4 shows the resource utilization summary for the entire BCNN implementation. LUTs are used for mapping all the computing kernels, including binary convolution, MP and NB kernels. Feature maps of convolutional layers are mapped onto distributed RAMs result in additional LUT consumption. The BRAM usage is mostly consumed by all the weight matrices. Flip-flops are used for storing feature maps and constructing a deep pipeline. Around 30% of the DSP slices are used by the $1^{st}$ layer to perform fixed-point multiplication. For the rest of convolutional layers, DSP slices are used for accumulating PE outputs as shown in Fig. 6.

Existing FPGA-based CNN implementations are compared in Table 5. To minimize the impact of different FPGA models on throughput, energy efficiency and performance density defined as throughput normalized to resource utilization are used as the performance metrics for comparison. Compared with the FPGA implementations of floating-point or reduced-precision CNNs, our BCNN implementation achieves 4-124x higher GOPS, 20-283x better energy-efficiency and 5-160x better performance density. Even compared with the BCNN implementation in Ref. 21, our work achieves 5x better performance density in terms of GOPS/kLUT. The work in Ref. 21 implements three kinds of computing kernels in hardware: floating-point convolution, binary convolution and fully-connected kernels. Since this reference work maps a single layer of the BCNN at a time, only one kind of computing kernels is active at a time. Such a time multiplexing scheme limits the system throughput due to the low hardware utilization. In our design, all the layers of the BCNN are mapped into a streaming architecture with optimized architectural parameters, and the data is flowing throughout the entire architecture in a deep pipeline. Therefore, the kernels are constantly active, and the utilization rate of the hardware resources is high. In addition, Ref. 21 consumes extra power for loading the weights from off-chip memory layer by layer in addition to the FPGA power reported. On the contrary, there is no such overhead in our architecture since we fully map the network and trained parameters on chip.

### 6.3 FPGA-based verse GPU-based BCNN

Fig. 7 compares the performance of the BCNN accelerated by a Titan X GPU and our FPGA-based design. For GPU acceleration, the baseline kernel is designed for floating-point computation, and the XNOR kernel is optimized for bitwise operations [9]. In the XNOR kernel, it concatenates 32 1-bit values into a 32-bit value. At the peak performance, each CUDA core can execute 32 bitwise operations per clock cycle. That is the reason why BCNN can also gain remarkable speedup on a GPU when using the XNOR kernel for compilation.

GPU acceleration is apparently sensitive to the size of workload (batch size here). One of the keys to achieving high performance in GPU computing is to hide the long latency of functional units by data-level interleaving especially when there are loop-carried data dependency existed in the algorithm. Only when the workload is large enough, a GPU is able to maintain high thread-level parallelism to achieve a high throughput. Differently, the FPGA-based solution is invariant to the batch size of data. Experiment results show that our design significantly outperforms the GPU acceleration using the baseline kernel in terms of both throughput and energy efficiency. Even compared with the GPU acceleration using the XNOR kernel, which is reported as the best GPU-based CNN performance by far, our design achieves a 75x better energy efficiency and an 8.3x better throughput for processing data in a small batch size of 16. For processing data in a large batch size of 512 (the maximum size that fit into the GPU memory), our design can match the throughput of the GPU acceleration with a 9.5x better energy efficiency.

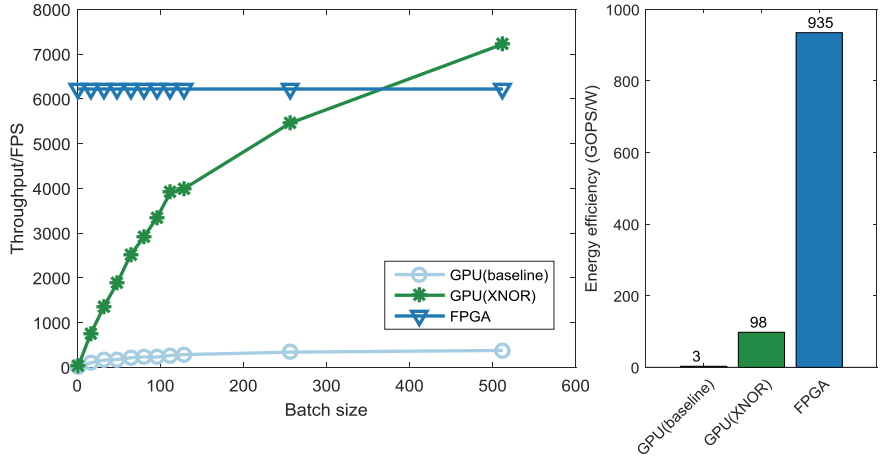

* Titan X GPU has used up ~80% memory for batch size of 512

Fig. 7. Throughput and energy efficiency comparison with GPU implementations.

Therefore, the FPGA-based BCNN solution is a clearly better choice for accelerating the data center applications that process online individual requests in small batch sizes. In a recent study conducted by Baidu, a dominant Internet company in China with 600 million active users, it is reported that the typical on-line prediction workload in terms of batch size is around 8 to 16 [25]. Such small workload is not enough for GPU to achieve its peak throughput performance. Thus, the FPGA-based solution is more superior in handling this kind of requests from individual users.

For processing static data in large batch sizes, the proposed solution is on a par with a Titan X GPU in terms of throughput while delivering much higher energy efficiency. This renders the FPGA-based solution a better choice for energy constrained applications, such as mobile-based advanced driver assistance systems (ADAS). In the ADAS application, a large batch of data needs to be processed for monitoring real-time road condition. In this case, both throughput and energy efficiency are essential and the FPGA-based solution can be deployed.

## 7 CONCLUSION

In this paper, we propose an optimized accelerator architecture tailored for BCNNs. We demonstrate for the $1^{st}$ time that the FPGA-based BCNN solution can greatly outperform a Titan X GPU in terms of both throughput and energy efficiency for processing accurate image classification tasks. The proposed BCNN accelerator running on a Virtex-7 FPGA is 8.3x faster and 75x more energy-efficient than a Titan X GPU for processing individual online requests in small batch sizes. For processing static data in large batch sizes, the proposed solution is on a par with a Titan X GPU in terms of throughput while delivering 9.5x higher energy efficiency. Thus, BCNNs are ideal for efficient hardware implementations on FPGAs regardless of the size of workload. The bitwise operations in BCNNs allow for the efficient hardware mapping of convolution kernels using LUTs, which is the key to enable massive computing parallelism on an FPGA. Applying the optimal levels of architectural unfolding, parallelism, and pipelining based on the proposed throughput model is the key to maximizing the system throughput.

Building memory channels across layers with data-flow control is the key to constructing a streaming architecture to further improve the throughput.

**ACKNOWLEDGMENTS**

This work by Arizona State University and Nanyang Technological University is supported by Cisco Research Center (CG#594589) and Singapore MOE Tier-2 (MOE2015-T2-2-013), respectively. We acknowledge Mr. Skip Booth and Mr. Hugo Latapie from Cisco for fruitful research discussions. We also thank Xilinx University Program for donating the FPGA boards.